\documentclass[aps,pre,reprint,groupedaddress,showpacs]{revtex4-1}
\usepackage{amsmath,amssymb,graphicx,color}

\newcommand{\EE}{\mathbb{E}}
\newcommand{\PP}{\mathbb{P}}
\newcommand{\la}{\langle}
\newcommand{\ra}{\rangle}

\begin{document}



\title{Memory and limit cycles in rock-scissors-paper}


\author{James Burridge}
\affiliation{Department of Mathematics, University of Portsmouth, UK}


\date{\today}

\begin{abstract}
When playing games in groups, it is an advantage for individuals to have accurate statistical information on the strategies of their opponents. Such information may be obtained by remembering previous interactions. We consider a rock-scissors-paper game in which agents are able to recall their last $m$ interactions, used to estimate the behaviour of their opponents. At critical memory length, a Hopf bifurcation leads to the formation of stable limit cycles. In a mixed population, agents with longer memories have an advantage, provided the system has a stable fixed point, and there is some asymmetry in the payoffs of the pure strategies. However, at a critical concentration of long memory agents, the appearance of limit cycles destroys their advantage. By introducing population dynamics that favours successful agents, we show that the system evolves toward the bifurcation point.
\end{abstract}

\pacs{02.50.Le,05.10.Ln}

\maketitle

\section{Introduction}

``The wise can learn from their enemies'' \cite{Aris}, and how they do this is an interesting question. The mathematical analysis of competitions between opponents, called ``the theory of games'', began with von Neumann \cite{Neu53} and Nash \cite{Nash51}. This early work concerned the equilibria of games, where no agent has anything to gain by changing strategy \cite{hof98}. The question of if, and how, game players reach an equilibrium has been the subject of a great deal of work since. The dynamics of games can have different driving mechanisms. In ``evolutionary game theory'' \cite{May82}, devised by John Maynard Smith in 1973, agents (living organisms) whose strategies are effective against competitors survive and reproduce more rapidly. This replicates their genes, and strategies they encode, in higher concentration in future generations. The ``replicator equations'' which describe this process have applications in diverse scientific settings from the evolution of language \cite{Mit04} to behavioural dynamics and decision making \cite{Pai12}. Alternatively, death and reproduction need not be involved if agents are able to learn from experience. An early example of such a learning rule is ``fictitious play'' \cite{Bro51} where players believe that their opponents are choosing strategies at random from a stationary distribution. They build a progressively clearer picture of this fictitious distribution through repeated interactions and at each round play the best response. The rule has an obvious flaw in that the real distribution is non-stationary. An alternative, derived from psychology, is reinforcement learning \cite{Sut98} of which an an example is ``experience weighted attraction'' \cite{Gal13}. Here, actions that have proved successful in the past are played more frequently. In the context of cyclic competition, this can give rise to a wide range of competitive and cooperative behaviours, including quasiperiodicity, limit cycles, intermittency and chaos \cite{Sat03}. If agents learn by sampling  a finite number of their opponents' moves between strategy updates, the noise inherent in these samples has been shown \cite{Gal09} to lead to noise-sustained cycling or removal of periodic orbits present in the limit of infinite sample size. The processes of sampling in between strategy updates is referred to as ``batch learning''.

In this paper we introduce learning dynamics in which agents possess a simple form of finite memory for their previous interactions. They use this to predict the current best pure strategy, adjusting their probabilistic strategy after each new interaction. Each agent's memory acts as a sample used for ``online learning'' as opposed to batch learning, and this distinction is central to the effects we uncover. We use our learning rule to investigate the children's game of rock-scissors-paper where rock blunts scissors, scissors cut paper, and paper wraps rock. The three strategies cyclically dominate each other \cite{Sza07}, a situation which can arise in nature. For example, male side-blotched lizards \cite{Sin96} adopt one of three mating strategies: \emph{ultra-dominant} with a large territory, \emph{mate guarding} in a small territory or \emph{sneaker} without territory, mating opportunistically. Dominant lizards beat guarders, but are vulnerable to sneakers, whereas guarders beat sneakers, creating a cyclical competition and oscillations in the frequencies of the three strategies. Cyclic competition can also arise in sociological contexts \cite{Sem03}. When the rock-scissors-paper game is studied using the replicator equations, the dynamics lack stable limit cycles. Depending on the values of payoffs for winning or losing, the system exhibits one of three kinds of behaviour: stable coexistence, neutrally stable cycles, or cycles of increasing amplitude \cite{Now06}. The first result we present is to show that in our memory based learning dynamics, limit cycles can form at critical memory length via a Hopf bifurcation \cite{Ern09}. The appearance of such cycles, also created by a Hopf bifurcation, has recently been discovered in the replicator equations, provided that mutations from one strategy to another \cite{Mob10}, or more complex patterns of mutation \cite{Tou15} are allowed. In contrast to this work, due to the memory present in our system, our dynamics is most naturally described by delay equations.

The neurological mechanisms by which humans and animals remember information are not completely understood \cite{Tra10}, but it is clear that there are multiple types of memory, and memory systems \cite{Mil98}. Broadly speaking, memories fall into two classes: they are either explicit recollections of events or facts - ``declarative'' memories - or learned motor or social skills and social conditioning \cite{Tra10}. Our agents are endowed with a simple, finite declarative memory for interactions, but our analysis could be repeated with other models of memory, or an empirical ``forgetting curve'' \cite{Ave10} which describes how memory decays with time. Our current aim is to investigate how the length of an agent's memory can influence their individual effectiveness, and the dynamics of the game. In common with earlier investigations of the use of sampling to determine strategy updates \cite{Gal09}, we find that sample size (memory length), which determines the strength of noise in the data, has a powerful effect. After demonstrating the appearance of limit cycles, we investigate how reduced noise aids decision making. Because each agent's memory includes both recent and older behaviour, then strategy updates are made based on slightly out-of-date information. Provided that strategy adjustments are sufficiently small or, equivalently, memory is not too long, then the game has a stable fixed point, and agents with long memories fare better than their short memory counterparts; their prediction of the best strategy is subject to smaller random errors. However, the limit cycles which appear when long memory agents are in high enough concentration destroy their competitive advantage.

\section{Model Definition}

We study the zero sum rock-scissors-paper game, with payoff matrix
\begin{equation}
\label{POM}
\bordermatrix{      &_R& _S& _P  \cr
                _R &0 & \alpha & -\gamma  \cr
                _S & -\alpha & 0 & \beta \cr
                _P &\gamma & -\beta & 0  \cr}.
\end{equation}
The game is played in continuous time by $L \ge 2$ agents who interact using random pairings which occur at rate $L/2$ per unit time so that each agent experiences, on average, one interaction per unit time. Formally, the probability that a single pairing takes place in time $\delta t$ is $L \delta t/2 + o(\delta t)$. Agents each adopt a probabilistic strategy which, for agent $i \in \{1,\ldots\L\}$, after the $n$th interaction is written $[r_i,s_i]_n$, where $r_i,s_i$ and $1-r_i-s_i$ are the probabilities of playing, respectively, rock, scissors or paper at the next interaction. Each agent is able to recall his last $m$ interactions, producing a sample from the population of strategies $\{R,S\}$ where $R$ and $S$ are the numbers of rock and scissors interactions in his memory. Note that at any given time, both samples $\{R,S\}$ and strategies $[r_i,s_i]_n$ require only two parameters for their description.

\begin{figure}
\includegraphics[width=6cm]{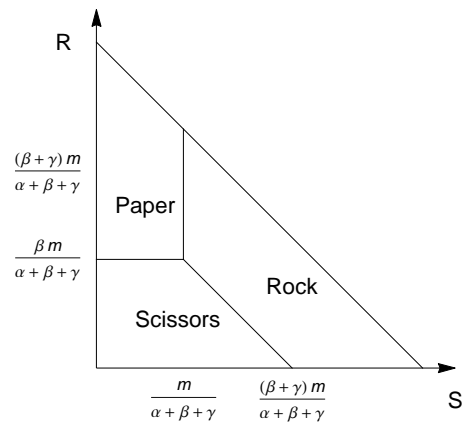}
\caption{Domains in which each of the three strategies are assessed to be the best.  \label{rspDoms}}
\end{figure}

Agents estimate the current average probability weights of their opponents as the fractions in their current sample, and use this to discern the optimal strategy. From the form of the payoff matrix (\ref{POM}), we see that the optimal strategy is determined by which of the following domains the current sample lies
\begin{align}
\label{RDom}
\mathcal{D}_{R} & := \left\{ S > \frac{\gamma m}{\alpha + \beta+ \gamma} , R > \frac{(\beta+\gamma)m}{\alpha + \beta + \gamma} - S \right\}  \\
\label{SDom}
\mathcal{D}_{S} & := \left\{ R < \frac{\beta m}{\alpha + \beta + \gamma} , R < \frac{(\beta+\gamma)m}{\alpha + \beta + \gamma} - S \right\} \\
\label{PDom}
\mathcal{D}_{P} & := \left\{ R > \frac{\beta m}{\alpha + \beta + \gamma} , S < \frac{\gamma m}{\alpha + \beta + \gamma} \right\}.
\end{align}
These domains are illustrated in Figure \ref{rspDoms}.
At each point in time, each agent's memory defines the position of a random walker in the simplex $\{(R,S) | R+S \leq m\}$, with each new interaction determining the next step of the walk. After each step an agent will update his strategy according to the domain in which its memory lies, using the following rule
\begin{equation}
[r_i,s_i]_{n+1} = (1-\epsilon) [r_i,s_i]_n + \epsilon
\begin{cases}
 [1,0] \text{ if } (R,S) \in \mathcal{D}_R \\
 [0,1] \text{ if } (R,S) \in \mathcal{D}_S \\
 [0,0] \text{ if } (R,S) \in \mathcal{D}_P.
\end{cases}
\label{EMA}
\end{equation}
For certain combinations of payoffs and memory length, $(R,S)$ can lie on the boundary between two domains, in which case no update is made. The parameter $\epsilon \in [0,1]$, the ``update rate'', describes the sensitivity of agents to new information, and $n$ indexes the number of interactions. According to definition (\ref{EMA}), the current strategy is an exponential moving average of the strategies which were estimated to be optimal from past interactions. Our learning rule has some commonality with experience weighted attraction rules used in recent studies \cite{Gal09,Gal13} where a parameter analogous with our $\epsilon$ is used to describe how rapidly agents respond to new information. In these studies, the parameter is interpreted as a measure of memory in the learning process, because it determines how much weight is given to previous information. Using this interpretation, our model may be seen as containing two kinds of memory: an ``active'' memory - the recent sample - used for decision making, and a ``passive'' memory which equally weights all previous samples in the limit $\epsilon \rightarrow 0$ and only uses the most recent when $\epsilon = 1$. We note also that our rule, in common with fictitious play \cite{Bro51}, implicitly assumes that agent's samples are drawn from a stationary distribution.

\section{Simulation Results}

\subsection{Single memory length}

\begin{figure}
\includegraphics[width=8cm]{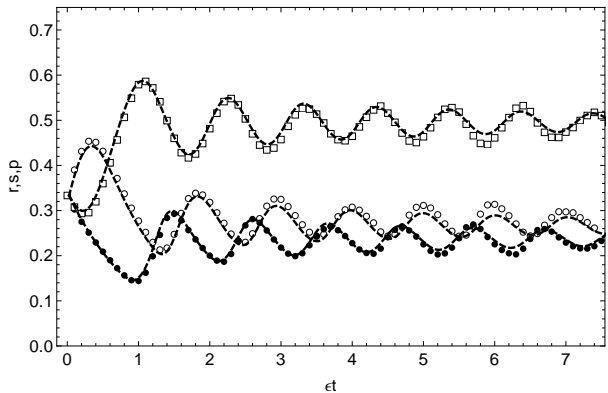}
\caption{ Probability weights of rock (open circles), scissors (dots) and paper (squares) agents in a group of size $L=100$ with $\alpha=2, \beta= \gamma=1$. All agents have memory $m=100$ and update rate $\epsilon = 5 \times 10^{-4}$. Dashed lines show solutions to delay equations (\ref{rDelay}) and (\ref{sDelay}) using the same parameter values. \label{stable}}
\end{figure}

\begin{figure}
\includegraphics[width=8cm]{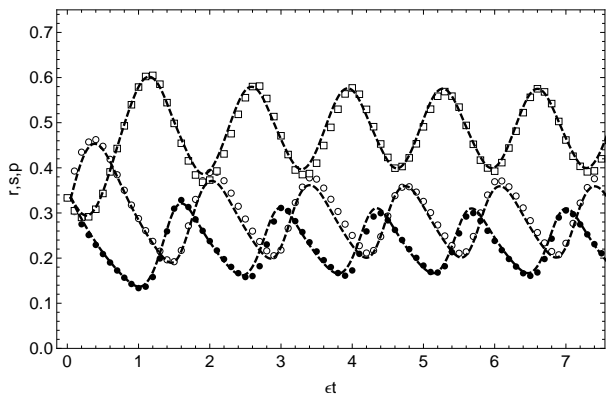}
\caption{Probability weights of rock (open circles), scissors (dots) and paper (squares) agents in a group of size $L=100$ with $\alpha=2, \beta=\gamma=1$. All agents have memory $m=100$ and update rate $\epsilon = 10^{-3}$. Dashed lines show solutions to delay equations (\ref{rDelay}) and (\ref{sDelay}) using the same parameter values. \label{hopf}}
\end{figure}

We begin by considering the behaviour of a population of agents, all of whom have the same memory length and update rate. In order to explore the dynamics of the population, we observe the average probability weights associated with each strategy. For example, the weight associated with rock is
\begin{equation}
r(t) := \frac{1}{L} \sum_{i=1}^L r_i(t)
\end{equation}
with $s(t)$ and $p(t)$ defined similarly. The evolution of the total probability weights in a population with update rate $\epsilon = 5 \times 10^{-4}$ and memory $m=100$ is shown in Figure \ref{stable}. Initial oscillations decay, eventually leaving the system in a stable state. The strategies of individuals in the population track the curves in Figure \ref{stable}, but with a greater stochastic component. In Figure \ref{hopf} we have increased the rate to $\epsilon=10^{-3}$, and we see that stable oscillations have formed. For given memory length, these appear when the update rate is sufficiently large. Conversely, for given update rate, stable oscillations emerge when the memory length is sufficiently large. We will show later that it is the product $\epsilon m^2$ which must exceed a critical value for stable oscillations to appear. This transition from stable equilibrium to stable oscillations at critical parameter values is known as a ``Hopf Bifurcation'' \cite{Ern09}.

\subsection{Agents as statisticians}

The memory of each agent represents a sample from a time varying population, which is used to estimate the current properties of the population. It is the fact that the sample is not drawn from the current population which allows oscillations to form. Since the primitive method used by our agents to estimate strategy fractions plays a role in destabilizing the fixed point, it is interesting to compare it to more sophisticated methods in order to determine its plausibility as the choice of an agent with some degree of ``common sense'', assuming he can recall the times at which the interactions in his sample took place. We do not formally address the question of which estimation technique would be selected by rational agents with unlimited intellectual resources.

In the same way that the method of regression is used to estimate, by maximum likelihood, the values of continuous explanatory variables given a sample of continuous response variables, so logistic regression \cite{Sha15} produces estimates when the response variables are discrete. The explanatory variables in our system are the collective time varying strategies of the group, whereas the response variables are the sequences of observations of strategy types that each agent makes.  In order to investigate how logistic regression compares against our agents' naive approach, as a technique for estimating the current strategy weights, we will use it to estimate the current rock weight given a finite memory sample.

Letting $t_{ik}$ be the time at which agent $i$ experiences his $k$th interaction back from the current time, and $I_R(i,k)$ be the indicator function that this interaction is with a rock agent, then the likelihood of his current memory will be
\begin{equation}
\mathcal{L}_i := \prod_{k=1}^{m} r(t_{ik})^{I_R(i,k)}[1-r(t_{ik})]^{1-I_R(i,k)}.
\end{equation}
We now suppose that during the period of time covered by the agent's memory, the rock probability weight may be expressed as a logistic function
\begin{equation}
\label{rlogit}
r(t) = \frac{1}{1+e^{-(\beta_0 + \beta_1 t)}}
\end{equation}
where $\beta_0$ and $\beta_1$ are constants. It is straightforward to carry out higher order regression where the exponent in the denominator is replaced with a polynomial of higher order. However, the length of time covered by agents' memory will typically be significantly shorter than the period of oscillation, so that $r(t)$ changes in an approximately linear way whilst their samples are being collected. For example the period of oscillation in Figure \ref{hopf} is $T \approx 1375$ whereas the memory length is $m=100$, and the greatest change in $r(t)$ during a single time interval of length 100 is 0.016.  We later demonstrate that the ratio of the period of stable oscillations to memory length obeys
\begin{equation}
\frac{T}{m} \propto \sqrt{m},
\end{equation}
at the point where oscillations begin to form, so that in systems where agents have longer memories, these memories cover a relatively shorter fraction of one oscillation.

To determine the values of $\beta_0, \beta_1$ which maximize the likelihood $\mathcal{L}_i$ we express the log likelihood as a function of these parameters:
\begin{align}
\ln \mathcal{L}_i  &= \sum_{k = 1}^{m} I_R(i,k)[\beta_0 + \beta_1 t_{ik}] \\ &- \sum_{k = 1}^{m}\ln(1+e^{\beta_0 + \beta_1 t_{ik}}).
\end{align}
This expression is then numerically maximized to find the most likely values of $\beta_0$ and $\beta_1$ which can then be used in equation (\ref{rlogit}) to predict the current value of $r(t)$.
\begin{figure}
\includegraphics[width=8cm]{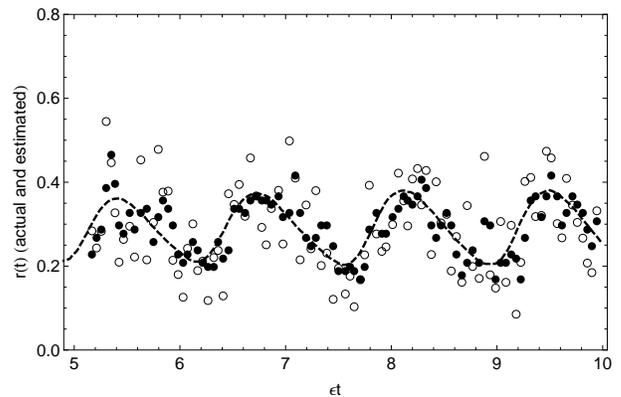}
\caption{ Dashed line shows the rock weight in a group of size $L=100$ with $\alpha=2, \beta=\gamma=1$. All agents have memory $m=100$ and update rate $\epsilon = 10^{-3}$. Open circles show binomial logistic regression estimates of the rock fraction made using the memory of a single randomly selected agent at a sequence of regularly spaced intervals. Black dots show similar estimates made by simply computing the fraction of rock interactions in the agent's memory.
\label{logit_mean}}
\end{figure}
In Figure \ref{logit_mean} we see the results of this method, applied to the memory of a single randomly chosen agent from the simulation in Figure \ref{hopf}. Also shown are the results of the primitive estimation method used by our agents. The primitive method appears considerably more effective at accurately predicting the current weights. However, a subtle phase shift may be perceived in the agent's predictions relative to the true values. In order to investigate this shift, and to compare the two methods in greater detail we construct a fictitious sequence of Bernoulli random variables $\{X_k\}_{k=1}^{m}$  with success probability, $p(k)$, which changes over the course of the sequence: $p(k)=p_0 + (p_1-p_0)k/m$. We then use our two methods to find an estimate $\hat{p}_1$ of the value of $p_1$, for a very large number of example sequences, allowing us to estimate the distribution of $\hat{p}_1$.
\begin{figure}
\includegraphics[width=8cm]{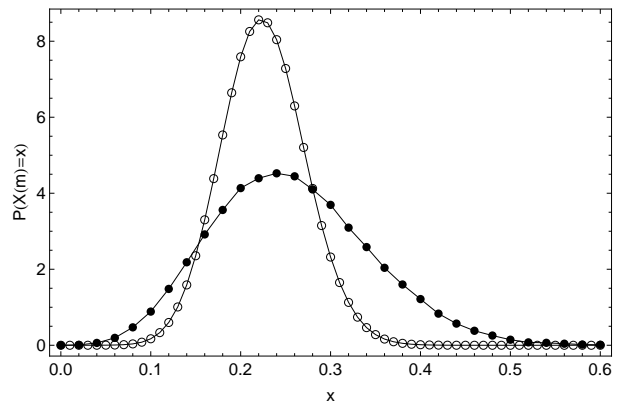}
\caption{ Given a sequence $\{X_k\}_{k=1}^{m}$ where $m=100$, of Bernoulli random variables with changing success probability $p(k)=p_0 + (p_1-p_0)k/m$ where $p_0=0.2, p_1=0.25$. Black dots show the estimated probability density function of the logistic regression estimator $\hat{p}_1$, having mean 0.256 and root mean squared error $0.07$. Open circles show the estimated probability distribution of the number of successes in the sequence, serving as an alternative estimator for $p_1$, having mean 0.225 and root mean squared error 0.07. \label{logit_dist_100}}
\end{figure}
\begin{figure}
\includegraphics[width=8cm]{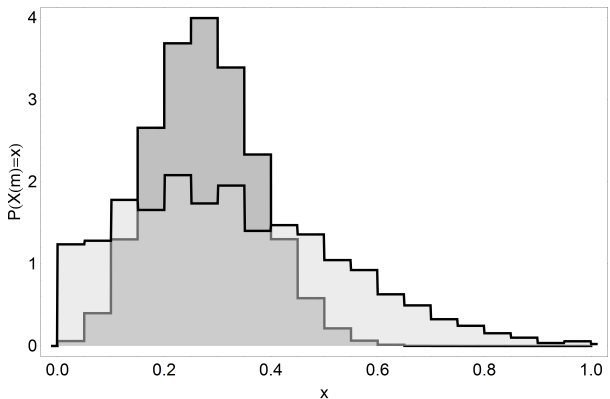}
\caption{ Given a sequence $\{X_k\}_{k=1}^{m}$ where $m=20$, of Bernoulli random variables with changing success probability $p(k)=p_0 + (p_1-p_0)k/m$ where $p_0=0.2, p_1=0.3$.  Light gray histogram gives estimated probability distribution of the logistic regression estimator $\hat{p}_1$, having mean 0.33 and root mean squared error $0.2$. Dark gray histogram gives estimated probability distribution of the number of successes in the sequence, having mean 0.25 and root mean squared error $0.1$. \label{logit_dist_20}}
\end{figure}
In Figures \ref{logit_dist_100} and \ref{logit_dist_20} we have constructed distributions for $\hat{p}_1$ in cases $m=100$ and $m=20$. It is clear that logistic regression produces estimates with much higher variance, but with smaller bias. Whilst the primitive method is the more efficient estimator, in the sense that the mean squared error in its predictions is smaller, its bias creates a systematic delay in its predictions. For larger memory values this bias will reduce, because the period of oscillation grows faster than the memory length, so the weights change by a smaller amount over the course of each agent's memory.

\subsection{Mixed memory and population dynamics}

\begin{figure}
\includegraphics[width=8cm]{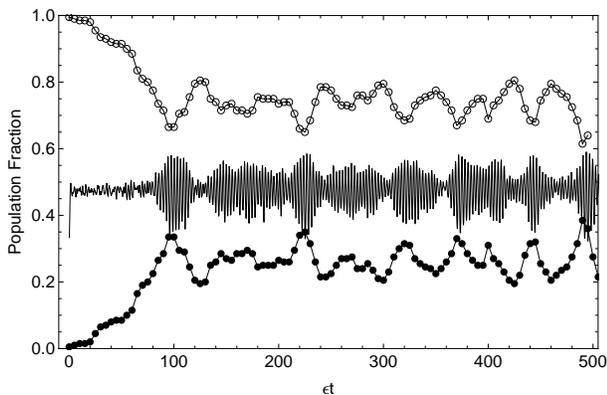}
\caption{Populations of $m=10$ (open  circles) and $m=100$ (dots) agents in system of size $L=200$ with update rate $\epsilon = 0.01$ initially composed of 199 short memory agents and a single long memory agent. Game parameters are $\alpha = 2, \beta = \gamma =  1$ and population dynamics parameters are $\rho = \kappa = 10^{-3}$. Also shown in fraction of paper agents (black line), illustrating how the onset of oscillations coincides with stable equilibrium between memory lengths.  \label{popDyn}}
\end{figure}

We now consider the case of a mixed population containing two memory lengths, $m \in \{10,100\}$. Intuitively we expect that agents with longer memories will be at an advantage because they are able to more accurately predict the optimal strategy; their estimates of opponents' strategies are subject to smaller errors. To investigate this we introduce some simple population dynamics, based upon an exponential moving average of each agent's payoffs. We let $v_{i k}$ be the payoff to agent $i$ at his $k$th interaction since the start of the game, and define his moving average, $\bar{v}_{ik}$, as follows
\begin{equation}
\bar{v}_{i,k+1} := (1-\rho) \bar{v}_{ik} + \rho v_{i,k+1}.
\end{equation}
At each pairwise interaction in the game, with probability $\kappa \ll 1$, the agent with the lower average payoff is replaced with a copy of the higher scoring agent and the total score of the pair shared equally between the original and its copy. In this way the total payoff in the population remains at zero. In Figure \ref{popDyn} we show the evolution of the fractions of short and long memory agents in such a simulation, along with the probability weight for paper, as a signifier for the presence of oscillations. Initially we have a single long memory agent, whose descendants reproduce rapidly due to their enhanced ability to determine the optimal strategy to play. However, once these long memory agents are in sufficient concentration, they create a limit cycle. The presence of this cycle gives short memory agents an advantage, because although their samples have a higher noise, they contain more recent data. In circumstances of rapid change, older data becomes irrelevant, and skews the sample of the long memory agents, leading them to make poorer choices. In consequence the effectiveness of the two memory lengths comes into balance. The population sizes undergo noisy oscillations about the bifurcation point once their collective strategies begin to oscillate. This indicates that the bifurcation point is itself a self organized state, driven by population dynamics.

The fact that the payoffs of the three pure strategies are not equal is essential in order for long memory players to have an advantage. Simulations similar to that illustrated in Figure \ref{popDyn} show that as the payoffs $\alpha, \beta, \gamma$ approach equality, the long memory players do not prosper and the self organizing Hopf bifurcation is no longer present. An intuitive understanding of this may be obtained by considering the effect of shortening the memory length, which causes individual agents' samples to perform higher variance random walks around the simplex of Figure \ref{rspDoms}. As the variance of the walk increases, agents' predictions of the best strategy are subject to greater variability, driving their strategies toward the symmetric case $[\tfrac{1}{3},\tfrac{1}{3}]$. If the game is not symmetric then a small shift towards the symmetric strategy for short memory players gives an advantage for longer memory players. In the symmetric case, this advantage disappears because the symmetric strategy is optimal.

\section{Theory}

Here we investigate the dynamics of a system of identical agents as $m \rightarrow \infty$ and $L \rightarrow \infty$. The results we obtain by considering these limits provide excellent approximations to the behaviour of smaller groups of agents for a wide range of memory values. In particular we study the symmetric game, allowing us to discover the analytical conditions for stability and the period of oscillation.

\subsection{Individual estimates of optimal strategy}

In order to construct equations which describe the dynamics of the strategy weights $r(t), s(t), p(t)$, we require expressions for the probabilities that randomly selected agents will perceive each of the three strategies as optimal. If the strategy weights are constant then the numbers $R,S$ and $P=m-R-S$ of strategy types in a sample of size $m$ will be trinomially distributed with probability mass function
\begin{equation}
\label{tri}
f(x,y;r,s)= \frac{m! r^x s^y (1-r-s)^{m-x-y}}{x! y! (m-x-y)!}.
\end{equation}
In reality, the weights will change while the sample is being taken, producing a sequence of non-identical trivariate Bernoulli trials. Our aim is to approximate this non-stationary distribution with the stationary version (\ref{tri}) having appropriately chosen values of $r$ and $s$. We will discover, once our analysis is complete, that the period, $T$, of oscillations at the transition to instability satisfies $T \propto m^{3/2}$. Letting the magnitude of the change in $r(t)$ over a typical agent's sample be $\delta r$  then we see that $ \delta r = \mathcal{O} (m^{-1/2})$ as $m \rightarrow \infty$, with the same behaviour holding for $s(t)$. Thus the approximation of the sample distribution with a stationary distribution (\ref{tri}), becomes increasingly accurate as $m$ becomes large and our approximation is self consistent.

To calculate the probabilities that each of the strategies $\omega \in \{\text{Rock, Scissors, Paper}\}$ will appear optimal to a randomly selected agent, given recent behaviour of $r(t)$ and $s(t)$ we define indicator functions for the domains defined in (\ref{RDom}), (\ref{SDom}) and (\ref{PDom})
\begin{equation}
I_\omega (R,S) =
\begin{cases}
1 & \text{ if } (R,S) \in \mathcal{D}_\omega\\
0 & \text{ otherwise. }
\end{cases}
\end{equation}
In the limit $L \rightarrow \infty$ the interactions of an individual agent have negligible effect on the weights $r(t),s(t)$, which evolve deterministically. If we select an agent at random at time $t$, then conditional on the history of the weights $\{r(u),s(u)|u \leq t\}$, the probability that this agent will perceive $\omega$ as optimal is a functional of this history
\begin{equation}
\label{pomega}
\tilde{p}_\omega[r,s](t) := \EE[I_\omega(R,S)](t).
\end{equation}
Here the expectation is taken over the set of interaction times and types in the agent's memory. Letting $\tau_k$ be the time elapsed since the $k$th interaction back from time $t$, we define the following average
\begin{equation}
\la r \ra_m := \frac{1}{m} \sum_{k=1}^{m} r(t-\tau_k),
\end{equation}
which is a random variable depending on $\{ \tau_k\}_{k=1}^m$. Conditional on the interaction times, we approximate the non stationary distribution of types with the stationary distribution $f(x,y;\la r \ra_m, \la s \ra_m )$. We note that differences between second moments of this distribution and the true distribution will be of order $\delta r$ and $\delta s$, and the means will be identical. In order to average over the interaction times, we note that for a particular agent, the time intervals between interactions are exponentially distributed.  The distribution function for $\tau_k$ is therefore the Gamma density
\begin{equation}
\PP(\tau_k \in [t,t+\delta t]) = \frac{t^{k-1} e^{-t}}{(k-1)!} \delta t: = \gamma_k(t) \delta t
\end{equation}
so that
\begin{align}
\EE[\la r \ra_m] &= \frac{1}{m} \sum_{k=1}^m \int_0^\infty \gamma_k(u) r(t-u) du \\
&= \int_0^\infty \left(\sum_{k=1}^m \frac{\gamma_k(u)}{m} \right) r(t-u) du \\
&= \int_0^\infty \frac{\Gamma(m,u)}{m!} r(t-u) du \\
\label{gdef}
&:= \int_0^\infty g_m(u) r(t-u) ds
\end{align}
where $\Gamma$ is the incomplete gamma function, defined as
\begin{equation}
\Gamma(m,u) = \int_u^\infty x^{m-1} e^{-x} dx.
\end{equation}
Equation (\ref{gdef}) defines a probability density $g_m(s)$ having the shape of a smoothed top hat (uniform) distribution on $[0,m]$, and may be thought of as representing the strength of the collective memory of all agents $u$ time units before the present. As $m \rightarrow \infty$, the distribution of $\la r \ra_m$ becomes increasingly sharply peaked about its mean so
\begin{equation}
\EE[f(x,y;\la r \ra_m, \la s \ra_m )] \sim f(x,y;\EE[\la r \ra_m],\EE[ \la s \ra_m] )
\end{equation}
where the expectation is taken over interaction times. Defining $\bar{r}_m := \EE[\la r \ra_m]$ then:
\begin{align}
\label{pfunc1}
\tilde{p}_\omega[r,s](t) & \sim \sum_{x=0}^m \sum_{y=0}^m f(x,y;\bar{r}_m,\bar{s}_m) I_\omega(x,y) \\
&:= p_\omega(\bar{r}_m, \bar{s}_m)
\label{pfunc2}
\end{align}
as $m \rightarrow \infty$. Notice that the quantity $p_\omega(\bar{r}_m, \bar{s}_m)$ is a functional only because the time averages $\bar{r}_m, \bar{s}_m$ are functionals, whereas $p_\omega(\cdot,\cdot)$ is an ordinary function.

\subsection{Delay equation}
During a short finite time interval $[t,t+\delta t]$, given the history $\{r(u), s(u) | u \leq t\}$, from the learning rule (\ref{EMA}) the expected changes in weights will be
\begin{align}
\EE[\delta r] &= \epsilon [p_R(\bar{r}_m, \bar{s}_m) -r(t)] \delta t \\
\EE[\delta s] &= \epsilon [p_S(\bar{r}_m, \bar{s}_m) -s(t)] \delta t.
\end{align}
As $L \rightarrow \infty$, these changes become deterministic and the weights obey the following coupled delay equations
\begin{align}
\label{rDelay}
\frac{dr}{dt} &= \epsilon[p_R(\bar{r}_m, \bar{s}_m) - r] \\
\frac{ds}{dt} &= \epsilon[p_S(\bar{r}_m, \bar{s}_m) - s].
\label{sDelay}
\end{align}
Solutions to these equations are shown in Figures \ref{stable} and \ref{hopf}, where we see that they accurately capture the simulated evolution of a system with the same parameter values.

\subsection{Linear Stability Analysis for Symmetric Case}

In order to understand the conditions under which stable oscillations form, we examine the stability of the fixed point of equations (\ref{rDelay}) and (\ref{sDelay}). By considering the system in the symmetric case where $\alpha=\beta=\gamma=1$, we can derive analytical results which provide a qualitative understanding of the system in general. The fixed point of the system in the symmetric case is $(r,s)=(\tfrac{1}{3},\tfrac{1}{3})$.

\begin{figure}
\includegraphics[width=8cm]{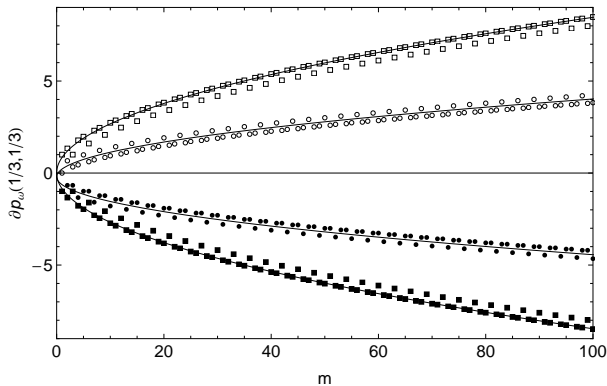}
\caption{The derivatives $\frac{\partial p_R}{\partial r}$, $\frac{\partial p_S}{\partial s}$, $\frac{\partial p_R}{\partial s}$, $\frac{\partial p_S}{\partial r}$ evaluated directly from equation (\ref{pfunc1}) at $(\bar{r}_m,\bar{s}_m)=(\tfrac{1}{3},\tfrac{1}{3})$. The four derivatives are plotted using open circles, filled circles, open squares, filled squares respectively. Black lines show corresponding expansion coefficients from equations (\ref{pRLin}) and (\ref{pSLin}). \label{drpR}}
\end{figure}

As $m \rightarrow \infty$, the trinomial distribution (\ref{tri}) becomes increasingly well approximated by a bivariate normal distribution \cite{Gri01} having the same means $\mu_R = m r, \mu_S = m s$, variances $\sigma_R^2 = m r(1-r), \sigma_S^2 = m s(1-s) $ and correlation $\rho = - (m r s)/(\sigma_R \sigma_S)$. By defining the parameter,
\begin{equation}
z = \frac{(x-\mu_R)^2}{\sigma_R^2} + \frac{2 \rho (x-\mu_R)(y-\mu_S)}{\sigma_R \sigma_S} + \frac{(y-\mu_S)^2}{\sigma_S^2}
\end{equation}
we can write this bivariate normal density as:
\begin{equation}
\label{bivnorm}
\phi(x,y)dx dy = \frac{\exp \left[-\frac{z(x,y)}{2(1-\rho^2)}\right]}{2\pi \sigma_R \sigma_S \sqrt{1-\rho^2}} dx dy.
\end{equation}
We can derive analytical approximations for the functionals $p_R(\bar{r}_m, \bar{s}_m)$ and $p_R(\bar{r}_m, \bar{s}_m)$, defined in (\ref{pfunc1}) and (\ref{pfunc2}), by integrating this density over the domains $\mathcal{D}_\omega$.  In the symmetric case this task may be achieved by introducing new variables \cite{Tou15} $(X,Y)$, related to $(x,y)$ as follows
\begin{align}
x &=  \frac{m}{3} + X - \frac{Y}{\sqrt{3}} \\
y  &=  \frac{m}{3} + \frac{2Y}{\sqrt{3}}.
\end{align}
We write the transformed density $\tilde{\phi}(X,Y;r,s)$ so that $\phi(x,y) dx dy = \tilde{\phi}(X,Y;r,s) dX dY$. Integrals of the density over the domains are tractable if the population fractions are $(r,s)=(\tfrac{1}{3},\tfrac{1}{3})$ so that the peak of the bivariate normal coincides with the vertex where the domains meet (see Figure \ref{rspDoms}), which is also the fixed point of the system in the symmetric case. We can exploit this tractability by expressing  $\tilde{\phi}(X,Y;r,s)$ as a perturbation of $\tilde{\phi}(X,Y;\tfrac{1}{3},\tfrac{1}{3})$. We first write the population fractions as perturbations about the fixed point
\begin{align}
r &= \frac{1}{3} + \psi_r \\
s &= \frac{1}{3} + \psi_s
\end{align}
where $\psi_r$ and $\psi_s$ are small fluctuations. We then define the following ratio
\begin{equation}
h(X,Y;\psi_r,\psi_s) := \frac{\tilde{\phi}(X,Y;\tfrac{1}{3}+\psi_r,\tfrac{1}{3}+\psi_s)}{\tilde{\phi}(X,Y;\tfrac{1}{3},\tfrac{1}{3})}
\end{equation}
which tends to unity as $\psi_r, \psi_s \rightarrow 0$. The first Taylor polynomial of $h$ about $\psi_r = \psi_s=0$ is
\begin{multline}
h_1(X,Y;\psi_r,\psi_s)  = 1 + \frac{6 X \left(m-\sqrt{3} Y\right)}{m} \psi_r \\
 + \frac{6 m \left(X+\sqrt{3} Y\right)-9 X^2-6 \sqrt{3} X Y+9 Y^2}{2 m} \psi_s.
\end{multline}
The density $\tilde{\phi}(X,Y;r,s)$ therefore has the following asymptotic behaviour, as $\psi_s, \psi_r \rightarrow 0$
\begin{equation}
\tilde{\phi}(X,Y;r,s) \sim h_1(X,Y;\psi_r,\psi_s) \frac{3 e^{-\frac{3(X^2+Y^2)}{m}}}{m \pi}.
\end{equation}
We now transform to polar coordinates $X=u \cos \theta, Y= u \sin \theta$, and make use of the fact that in these coordinates the domains are symmetrical with angular width $\tfrac{2\pi}{3}$. For example  $\mathcal{D}_R = \{ (u,\theta) | \theta \in [0,2\pi/3]\}$.
By integrating over these domains and defining time averaged fluctuations:
\begin{equation}
\bar{\psi}_r(t) := \int_0^\infty g_m(u) \psi_r(t-u)
\end{equation}
we obtain the following linearized expressions for $p_R$ and $p_S$:
\begin{align}
\label{pRLin}
p_R(\bar{\psi}_r, \bar{\psi}_s) &= \frac{1}{3} +  \frac{3(2 \sqrt{m \pi} - \sqrt{3})}{8\pi}\bar{\psi}_r  + \frac{3 \sqrt{m}}{2 \sqrt{\pi}} \bar{\psi}_s\\
p_S(\bar{\psi}_r,\bar{\psi}_s) &= \frac{1}{3} - \frac{3 \sqrt{m}}{2 \sqrt{\pi}} \bar{\psi}_r - \frac{3(2 \sqrt{m \pi} +\sqrt{3})}{8 \pi} \bar{\psi}_s.
\label{pSLin}
\end{align}
We can verify the quality of these approximations by comparing the expansion coefficients to the exact values of the derivatives of $p_\omega(r,s)$ evaluated at $(r,s)=(\tfrac{1}{3},\tfrac{1}{3})$, as shown in  Figure \ref{drpR}.
The linearized delay equations (\ref{rDelay}) and (\ref{sDelay}) may then be expressed in terms of the expansion coefficients:
\begin{align}
a(m) &:= \frac{3(2 \sqrt{m \pi} - \sqrt{3})}{8\pi}\\
b(m) &:= \frac{3 \sqrt{m}}{2 \sqrt{\pi}}  \\
c(m) &:= \frac{3(2 \sqrt{m \pi} +\sqrt{3})}{8 \pi}
\end{align}
as follows
\begin{align}
\label{linr}
\frac{\dot{\psi}_r(t)}{\epsilon} &= \int_0^\infty \left[a \psi_r(t-u) + b \psi_s(t-u)\right] g_m(u) du - \psi_r \\
\frac{\dot{\psi}_s(t)}{\epsilon} &= - \int_0^\infty \left[b \psi_r(t-u) + c \psi_s(t-u)\right] g_m(u) du - \psi_s
\label{lins}
\end{align}
As $m \rightarrow \infty$ the weight function $g_m(u)$ approaches a step function
\begin{equation}
g_m(u) \rightarrow
\begin{cases}
\frac{1}{m} \text{ if } u \in [0,m] \\
0 \text{ otherwise }
\end{cases}
\end{equation}
and the ratios $\tfrac{a(m)}{2 b(m)}, \tfrac{c(m)}{2b(m)} \rightarrow 1 + \mathcal{O}(m^{-1/2})$ so equations (\ref{linr}) and (\ref{lins}) approach the following simplified form:
\begin{align}
\label{linrSimp}
\frac{\dot{\psi}_r(t)}{\epsilon} &= \frac{3 }{4 \sqrt{m \pi}} \int_{t-m}^t \left[\psi_r(u) + 2 \psi_s(u)\right] du - \psi_r \\
\frac{\dot{\psi}_s(t)}{\epsilon} &= \frac{-3 }{4 \sqrt{m \pi}}\int_{t-m}^t \left[2\psi_r(u) + \psi_s(u)\right] du - \psi_s.
\label{linsSimp}
\end{align}
The symmetry of the system, together with numerical solutions to the full equations, suggests we search for oscillatory solutions which differ in phase by $2\pi/3$. We therefore make the anstaz $\psi_r(t) = e^{\lambda t}$ and $\psi_s(t) = e^{\frac{2\pi i}{3}} \psi_r(t)$, where $\lambda$ is a complex number. Substitution of these trial solutions into (\ref{linrSimp}) and (\ref{linsSimp}) yields two identical characteristic equations
\begin{equation}
\lambda^2 + \epsilon \lambda - \frac{3 \sqrt{3} \epsilon i}{4 \sqrt{\pi m}}(1-e^{-\lambda m}) =0.
\end{equation}
The fact that both characteristic equations are identical justifies our ansatz that pairs of solutions exist which are identical up to a phase shift. We now write $\lambda = x + i y$ and introduce the constant
\begin{equation}
A := \frac{3}{4} \sqrt{\frac{3}{\pi}}.
\end{equation}
The real and imaginary parts of the characteristic equation may then be written
\begin{align}
\label{re}
x^2-y^2 + \epsilon x + \frac{A \epsilon}{\sqrt{m}} e^{- m x} \sin(m y) &= 0\\
2 x y + \epsilon y - \frac{A \epsilon}{\sqrt{m}} [1- e^{-m x} \cos (m y)] &= 0.
\label{im}
\end{align}
For given memory length $m$, provided that $\epsilon$ is sufficiently small, the real part of the solutions to (\ref{re}) and (\ref{im}) will be negative, so the fixed point is stable. As we increase $\epsilon$, then $\lambda$ crosses through the imaginary axis, creating a switch to instability with oscillations of exponentially increasing magnitude. Although the fixed point of the full dynamics shares this transition to instability, the resulting oscillations are bounded, creating a stable limit cycle, the appearance of which is termed a ``Hopf Bifurcation'' \cite{Ern09}. To compute the critical value of the update rate $\epsilon_c$ we set $x=0$ in equation (\ref{im}) obtaining
\begin{equation}
\frac{1-\cos(m y)}{my} = \frac{1}{A \sqrt{m}}.
\end{equation}
This equation may have multiple roots, corresponding to different frequencies of oscillation. We may determine the asymptotic behaviour as $m \rightarrow \infty$ of the lowest root be expanding the left hand side to linear order about $y=0$, and then solving for $y$, obtaining
\begin{equation}
y \sim \frac{8}{3} \sqrt{\frac{\pi}{3}}m^{-\frac{3}{2}} \text{ as } m \rightarrow \infty.
\end{equation}
Substitution of this result into (\ref{re}) again with $x=0$ yields
\begin{equation}
\epsilon_c = \frac{256 \pi ^{3/2} \csc \left(\frac{8 \sqrt{\frac{\pi }{3}}}{3
   \sqrt{m}}\right)}{81 \sqrt{3} m^{5/2}} \sim \frac{32 \pi}{27 m^2} \text{ as } m \rightarrow \infty.
\end{equation}
From this analysis we see that the stability of the fixed point in the symmetric case depends of the value of the product $\epsilon m^2$. Our asymptotic expression for $y$ gives us the period of oscillation at the Hopf bifurcation point
\begin{equation}
T = \frac{3}{4} \sqrt{3 \pi m^3}.
\end{equation}
We now test these predictions numerically, and explore the non-symmetric case.

\begin{figure}
\includegraphics[width=8cm]{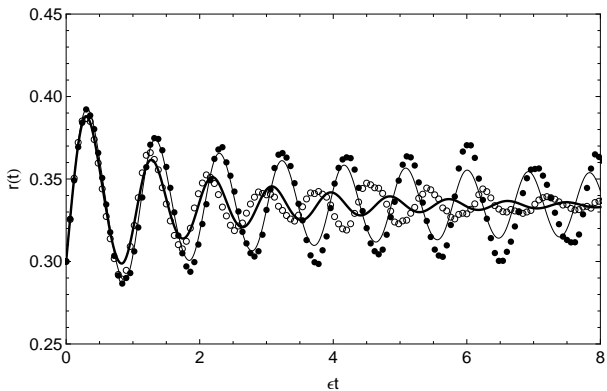}
\caption{Bold line shows $r(t)$ from the numerical solutions to equations (\ref{rDelay}) and (\ref{sDelay}) in the symmetric case when $m=100$ and $\epsilon=3 \times 10^{-4}$. Thin line show corresponding solution when $\epsilon=5 \times 10^{-4}$. Open circles and black dots show corresponding simulation results in a system of size $L = 100$. \label{hopfSym}}
\end{figure}


\begin{figure}
\includegraphics[width=8cm]{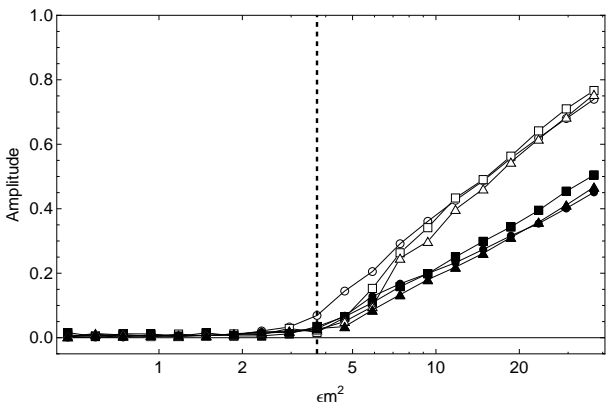}
\caption{Dependence of the steady state amplitude of $r(t)$ on $\epsilon$ for $m=20$ (open markers) and $m=80$ (filled markers). Three different combinations of payoff values $(\alpha, \beta, \gamma)$ were used: $(1,1,1)$ [circles],  $(1.5,2.5,1)$ [squares] and $(2,1,1)$ [triangles]. The vertical dotted line has horizontal coordinate $32 \pi/27$, corresponding to the critical value of $\epsilon m^2$ in the symmetric game. $L=500$ in all cases. \label{m_ep_hopf}}
\end{figure}

\subsection{Numerical tests of stability}

We consider the symmetric case first, both by numerically solving equations (\ref{rDelay}) and (\ref{sDelay}) where the probabilities $p_R$ and $p_S$ are as summations (\ref{pomega}) over the trinomial distribution, and by simulation. We consider the case $m=100$ and in Figure \ref{hopfSym} we have plotted $r(t)$ for values of $\epsilon$ lying just above and just below the critical value of $\epsilon_c = 2\pi / 16875 \approx 3.7 \times 10^{-4}$ predicted by our stability analysis. The appearance of stable oscillations is consistent with our analysis. We also numerically compute the period of oscillation at the critical point, finding that $T \approx 2347$ which compares to our analytical estimate of $T=740 \sqrt{3 \pi} = 2302$.

We now verify, using simulations, that the appearance of limit cycles depends on the value of the product $\epsilon m^2$ in three representative games, each with different payoffs. For two different memory values, $m \in \{20,80\}$ we have numerically determined the amplitude of oscillations in $r(t)$ for a series of values of $\epsilon$. These amplitudes are plotted in Figure \ref{m_ep_hopf} as functions of $\epsilon m^2$, and we see that the value $\epsilon m^2$ effectively predicts the onset of limit cycles at least at the levels of payoff asymmetry we have studied in this paper. We note however that we have found the precise critical value of this product only for the symmetric case in the limit of large $m$. From Figure \ref{m_ep_hopf} we observe that the asymmetric payoffs introduce small corrections to this critical value.

\section{Conclusion}

We have studied the rock-scissors-paper game played by agents with a simple form of memory. This memory is used by each agent to estimate the current best strategy. After each new interaction, agents incrementally update their own strategy, using a form of online learning. The naive technique for estimating strategy fractions used by our agents has, in common with fictitious play \cite{Bro51}, the underlying assumption of a stationary distribution of agent strategies in the population. Although this assumption is clearly false, we have shown that the technique can act as a more efficient estimator than logit regression.

Provided the system possesses a stable fixed point, agents with longer memories are able to more accurately determine the true weights, and therefore make better judgements about which strategy to play. However, excessively long agent memory produces a transition from stable equilibrium to a limit cycle. We have shown analytically in the symmetric case that the fixed point is destabilized when $\epsilon m^2$ reaches a critical value, and that the period of oscillations at the transition point grows as $m^{3/2}$. A simple form of population dynamics, imposed on a mixed population of long and short memory agents demonstrates that the initial advantage afforded long memory agents is destroyed when they become too numerous and destabilize the system.

Due to its role as the simplest model of cyclic competition, the rock-scissors-paper game is heavily studied. Recent work \cite{Mob10,Tou15} demonstrates how the introduction of mutations in to the replicator dynamics of the game can produce simple limit cycles via a Hopf bifurcation \cite{Ern09} in the replicator equations. We have shown that stable limit cycles can appear at critical agent memory, through a Hopf bifurcation in the delay equations which capture the learning dynamics of the population.

\bibliography{JBRefs}

\end{document}